\newif\ifdraft
\draftfalse
\newif\ifpreparepdf                       
\preparepdftrue 

        \ifdraft
\documentclass[aps,preprint,showpacs,longbibliography]{revtex4-1} 
        \else
\documentclass[aps,showpacs,superscriptaddress,longbibliography]{revtex4-1} 
        \fi

\ifpreparepdf
    \usepackage{color} 
    \usepackage[colorlinks]{hyperref} 
\else 
\fi

\usepackage{graphicx,amssymb,amsbsy}
\usepackage{ifthen}
\usepackage[percent]{overpic}
\usepackage{hyperref}
\usepackage{amsmath}

\graphicspath{{figs/}}  

\ifdraft    
   
   \newcommand{\PC}[1]{$\footnotemark\footnotetext{PC: #1}$}
   
   \newcommand{\NBB}[1]{$\footnotemark\footnotetext{NBB: #1}$}
   
   \newcommand{\APW}[1]{$\footnotemark\footnotetext{APW: #1}$}
   
   \newcommand{\KYS}[1]{$\footnotemark\footnotetext{KYS: #1}$}

   \newcommand{\file}[1]{$\footnotemark\footnotetext{{\bf file} #1}$}
   \newcommand{\mycomment}[2]{\noindent \textbf{\underline{#1}}: \emph{#2}}
   \newcommand{\edit}[1]{{\color{blue}#1}} 
\else   

   \newcommand{\PC}[1]{}
   \newcommand{\JFG}[1]{}
   \newcommand{\NBB}[1]{}
   \newcommand{\APW}[1]{}

   \newcommand{\KYS}[1]{}
   
   \newcommand{\file}[1]{}
   \newcommand{\mycomment}[2]{}
   \newcommand{\edit}[1]{{\color{black}#1}} 

\fi 

\newcommand{\arXiv}[1]{\href{http://arXiv.org/abs/#1}{arXiv:#1}}

\newcommand{\rf}     [1] {~\cite{#1}}

\newcommand{\refref} [1] {ref.~\cite{#1}}

\newcommand{\refeq}  [1] {(\ref{#1})}

\newcommand{\reffig} [1] {figure~\ref{#1}}

\newcommand{\beq}{\begin{equation}}

\newcommand{\eeq}{\end{equation}}
\newcommand{\ee}[1] {\label{#1} \end{equation}}

\newcommand{\bea}{\begin{eqnarray}}
\newcommand{\eea}{\end{eqnarray}}
\newcommand{\barr}{\begin{array}}
\newcommand{\earr}{\end{array}}

\newcommand{\etal}{{\em et al.}}    
\newcommand{\ie}{{i.e.}}        

\newcommand{\rpo}{rela\-ti\-ve periodic orbit}

\newcommand{\fFslice}{first Fourier mode slice}

\newcommand{\zeit}{\ensuremath{t}}  
\newcommand{\LieEl}{\ensuremath{g}}  
\newcommand{\id}{{\ \hbox{{\rm 1}\kern-.6em\hbox{\rm 1}}}}

\newcommand{\NSe}{Navier--Stokes equations}

\newcommand{\Reynolds}{\textit{Re}}  

\newcommand{\statesp}{state space}

\newcommand\Real{\mbox{Re}\,} 

\newcommand\flow[2]{{f^{#1}(#2)}}

\newcommand{\PoincM}{{\cal P}}      

\newcommand{\eigExp}[1][]{
\ifthenelse{\equal{#1}{}}{\ensuremath{\lambda}}{\ensuremath{\lambda^{(#1)}}}
                        }
\newcommand{\eigRe}[1][]{
\ifthenelse{\equal{#1}{}}{\ensuremath{\mu}}{\ensuremath{\mu^{(#1)}}}
                        }
\newcommand{\eigIm}[1][]{
  \ifthenelse{\equal{#1}{}}{\ensuremath{\omega}}{\ensuremath{\omega^{(#1)}}}
            }

\newcommand{\inprod}[2]{\left\langle #1 ,\, #2 \right\rangle}

\newcommand{\Diff}{\ensuremath{{\bf \textrm{D}}}}              

\begin{document}

\title{
Geometry of transient chaos in streamwise-localized pipe flow turbulence
}
\author{Nazmi Burak Budanur}
\affiliation{Nonlinear Dynamics and Turbulence Group,
             IST Austria,
             3400 Klosterneuburg, Austria}
\author{Akshunna Shaurya Dogra}
\affiliation{Department of Physics,
             New York University,
             726 Broadway, New York, NY 10003}
\author{Bj\"{o}rn Hof}
\affiliation{Nonlinear Dynamics and Turbulence Group,
             IST Austria,
             3400 Klosterneuburg, Austria}

\date{\today}

\begin{abstract}
In pipes and channels, the onset of turbulence is initially dominated by 
localized transients,
which lead to sustained turbulence through their collective 
dynamics.
In the present work, we study the localized turbulence in 
pipe flow numerically and
elucidate a state space structure that gives rise to transient chaos. 
Starting from the 
basin boundary separating laminar and turbulent flow, we identify transverse
homoclinic 
orbits, the presence of which necessitates a homoclinic tangle and chaos. 
A direct consequence of the homoclinic tangle is the fractal
nature of the laminar-turbulent boundary, which was conjectured
in various earlier studies. By mapping the transverse
intersections between the stable and unstable manifold of a periodic orbit, 
we identify the ‘gateways’ that promote an escape from turbulence.
\end{abstract}

\pacs{
02.20.-a, 05.45.-a, 05.45.Jn, 47.27.ed
            }

\maketitle

In wall-bounded shear flows close to onset, turbulence 
can be observed as localized patches that coexist with the orderly laminar 
flow that is stable against infinitesimal disturbances. 
In the case of pipe flow, these localized structures are known as puffs
and they have a simple phenomenology: After a chaotic time-evolution, 
a puff might suddenly decay and disappear or split and give rise to a new puff. 
As the governing parameter, the Reynolds number (\Reynolds), is increased, 
the mean puff lifetime before decaying 
increases while that before a splitting event decreases. 
This phenomenology was recently exploited to conceptualize a 
spatiotemporal description of the transition\rf{AMdABH11,MukHof2018}: 
In an infinitely long pipe that is initially populated 
with many puffs, turbulence will be sustained if the rate of puff splitting 
exceeds that of decay. 
Consequently, a critical $\Reynolds_C$ is defined as the $\Reynolds$, at 
which both rates are equal. 

In addition to allowing for a clear definition of the 
critical $\Reynolds_C$, the spatiotemporal dynamics also revealed
universal properties of the transition to turbulence: critical exponents 
were determined in laboratory experiments for Couette flow\rf{LSAJAH16}  and in 
direct numerical simulations for Couette\rf{LSAJAH16}  and 
Waleffe flow,\rf{ChTuBa17}\ 
and these suggest that the transition falls into the universality class of 
directed percolation.

Key to the above scenario is the existence of spatially independent
chaotic 
transients.
While the transient chaotic nature of the puff dynamics is evident 
from observations, how such dynamics arise from the Navier-Stokes equations 
is not well understood. 
Even though localized asymptotic states\rf{MMSE09} and 
invariant solutions\rf{ChWiKe14,AvMeRoHo13} at the laminar-turbulent boundary 
were computed, 
dynamical routes for sudden puff decay across the stable manifold of such 
solutions were not identified.
In this paper, 
we demonstrate the existence 
of homoclinic orbits in the vicinity of a periodic solution, 
which itself is on the laminar-turbulent boundary. 
Based on this observation, we suggest a global picture of the 
\statesp , containing a Smale horseshoe, which gives rise to  
transient chaos with a fractal basin boundary and thus, infinitely 
many paths for puff decay.

We simulate the incompressible fluid flow through a 
circular pipe of diameter $D$ and length 
$L = \pi D/ 0.1256637 \approx 25D$. For this purpose,
we employ \texttt{Openpipeflow}\rf{Willis2017}, which integrates
the perturbations 
${\bf u}(z, r, \theta; \zeit)$ to the base (Hagen-Poiseuille) solution 
${\bf u}_{HP}(z, r, \theta; \zeit) = 2 U (1 - (2r / D)^2) {\bf \hat{z}}$
of the pipe flow under periodic boundary condition 
${\bf u}(z, r, \theta; \zeit) = {\bf u}(z + L, r, \theta; \zeit)$
in the axial direction and no-slip boundary condition 
${\bf u}(z, r = D/2, \theta; \zeit) = 0$ at the pipe wall.
The Reynolds number is defined as $\Reynolds = U D / \nu$, 
where $\nu$ is the kinematic viscosity of the fluid. 
In our simulations, 
$\Reynolds = 1600$ and a constant flux is ensured via a
time-varying axial pressure gradient. 
For the numerical representation,
the flow fields are expanded in Fourier series truncated at 
$K = 255$ and $M = 47$ respectively in axial and azimuthal 
directions and with $64$ finite-difference points were used in 
the radial direction. 
The nonlinear terms are evaluated following the $3/2$-rule for 
dealiasing, leading to $(768 \times 144 \times 64)$ grid points 
in the physical space. 
    Time-stepping is performed by a second-order 
	  predictor-corrector method
	  with a time-step $\delta t \approx 0.00206 D/U$.
The adequacy of this resolution is confirmed in the previous studies
\rf{ChWiKe14,BudHof17}, which examined similar parameter 
regimes using \texttt{Openpipeflow} with slightly lower 
resolutions.
For the homoclinic orbits we are going to report in the following, the 
exponential drop-off of the Fourier coefficients is 4 orders of 
magnitude or more at every point along the orbit. 

As in the \refref{AvMeRoHo13}, the solutions that we consider are 
invariant under the reflection symmetry
and azimuthal rotation by $\pi$; 
\ie\ the flow fields satisfy 
$[u, v, w](z, r, \theta; \zeit) = [u, v, - w](z, r, - \theta; \zeit)$
and 
$[u, v, w](z, r, \theta; \zeit) = [u, v, w](z, r, \theta + \pi; \zeit)$, 
where $u,\,v$, and $w$ respectively denote the $z$, $r$, and $\theta$
components of the velocity field.
After these restrictions, the remaining symmetries of pipe flow
are the streamwise translations 
$\LieEl_z (l) {\bf u}(z, r, \theta) = {\bf u}(z - l, r, \theta)$
and the azimuthal rotation by $\pi/2$:
$\LieEl_{\theta} (\pi / 2) {\bf u}(z, r, \theta) 
 = {\bf u}(z, r, \theta - \pi / 2)$. Here, we are going to reduce
the translation symmetry via the \fFslice\ method \rf{BudCvi14}
following its adaptation to the pipe flow in
\refref{BudHof17}: Let 
$\hat{u}'_i
= J_0(\alpha r) \cos(2 \pi z/L)$, where $J_0$ is the 
Bessel function of the first kind, $\alpha$ is chosen such that
$J_0(\alpha 2r / D) = 0$, and $i = 1,2,3$	. Then the symmetry-reduced velocity fields 
are given by 
${\bf \hat{u}}(\zeit) = \LieEl_z (L \phi / 2 \pi) {\bf u}(\zeit)$,
where 
$	\phi (\zeit) = \arg (\inprod{{\bf u}(\zeit)}{{\bf \hat{u}'}} + i 
						 \inprod{{\bf u}(\zeit)}{\LieEl_z(- L/4) 
						 		 {\bf \hat{u}'}})$
and $\langle.,.\rangle$ denotes the $L_2$ inner-product.
See \refref{BudHof17} for details.
In the remainder of this paper we use the symmetry-reduced description and drop
the \^\ signs.

Avila \etal \rf{AvMeRoHo13} showed that in the above-described 
symmetry-invariant subspace, the laminar-turbulent boundary 
is set by the stable manifold of a 
relative periodic orbit, which is a periodic orbit with a 
streamwise shift. This orbit, which from here on we are going to 
refer to as LB, belongs to the lower branch of a pair of 
orbits that appear in a saddle-node bifurcation at 
$\Reynolds \approx 1450$ in our domain\footnote{
	The exact bifurcation point varies slightly depending  
	on the domain length and the resolution.}. 
After the symmetry-reduction, LB becomes
a periodic orbit, which satisfies 
${\bf u}_{LB} = \flow{T}{{\bf u}_{LB}}$, where 
$\flow{\zeit}{{\bf u (0)}} = {\bf u} (\zeit)$ denotes the 
finite-time flow mapping induced by the time-evolution under the
\NSe\ and the symmetry-reduction, 
$T$ is the period of LB, and ${\bf u}_{LB}$ is an arbitrary
point on the LB. Numerically, we converge to this solution
up to a relative error of $10^{-9}$ through
Newton-GMRES-hookstep iterations\rf{Visw07b} at 
$\Reynolds = 1600$ using $6000$ equally spaced time-steps along the 
orbit. We found LB's period at this \Reynolds\ to be 
$T \approx 12.35 \,D/U$ and in the subsequent computations we used a 
fixed time-step $\delta \zeit = T / 6000$\footnote{
This was necessary in order to compute the unstable manifold of  
LB accurately since the \rpo\ and its linear stability eigenvectors 
were converged with this time-step. We confirmed that this behavior is 
consistent by repeating parts of the computation with a time-step 
$\delta \zeit / 2$.}.

In order to study the linear stability and the unstable manifold of the LB, 
we define the Poincar\'e section as the set of velocity fields
${\bf \tilde{u}}$ that satisfy
\beq
	\inprod{{\bf \tilde{u}} - {\bf \tilde{u}}_{LB}}{
			 \partial_{\zeit} {\bf \tilde{u}}_{LB}} = 0
	\mbox{ and }
	\inprod{{\partial_{\zeit} \bf \tilde{u}}}{\partial_t {\bf \tilde{u}}_{LB}} > 0, 
	\label{e-pSect}
\eeq
where ${\bf \tilde{u}}_{LB}$ is an arbitrarily-chosen point on the LB.
Numerically, we approximated $\partial_{\zeit} {\bf u}$ as the first-order 
finite difference 
$\partial_{\zeit} {\bf u} \approx (\flow{\delta \zeit}{u} - u) / \delta \zeit$,
since its accuracy does not affect the accuracy of the steps that follow.
Geometrically, \refeq{e-pSect} describes a half-hyperplane in the 
infinite-dimensional \statesp\ that
accommodates the solutions to the \NSe . This construction, along 
with the evolution under the \NSe , implies a discrete-time dynamical 
system, \ie\ the Poincar\'e map
\beq
	{\bf \tilde{u}}[n + 1] = \PoincM ({\bf \tilde{u}}[n])
							 = \flow{\Delta \zeit_n}{{\bf \tilde{u}}[n]}\,,
	\label{e-pMap}
\eeq
where $n$ is the discrete-time variable and 
$\Delta \zeit_n$ is the first return time, which is the minimum time that 
is necessary for the \statesp\ trajectory to intersect the Poincar\'e 
section\refeq{e-pSect} after leaving it. 
	In order to find $\Delta \zeit_n$ in simulations, we first 
	integrate a trajectory with our fixed time-step $\delta \zeit$ and
	identify consecutive states for which the hyperplane condition 
	in \refeq{e-pSect} changes its sign from $-$ to $+$. We then 
	carry out a bisection search in time-steps until the hyperplane
    condition is satisfied with a relative error of $10^{-9}$.

\begin{figure}[h]
	\centering
	(a) \includegraphics[height=0.42\textwidth]{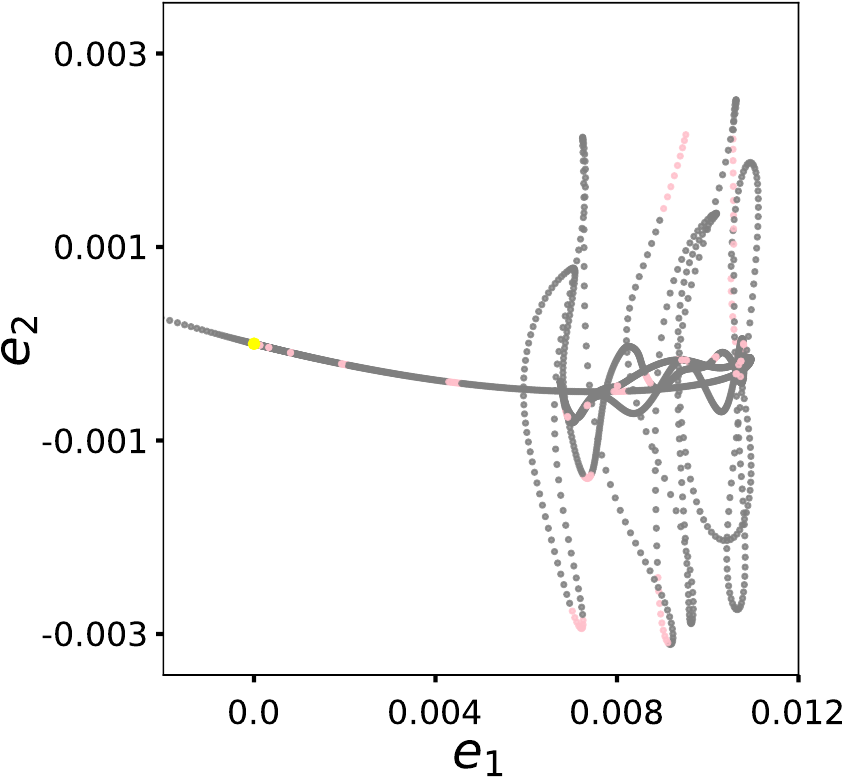} 
	(b) \includegraphics[height=0.42\textwidth]{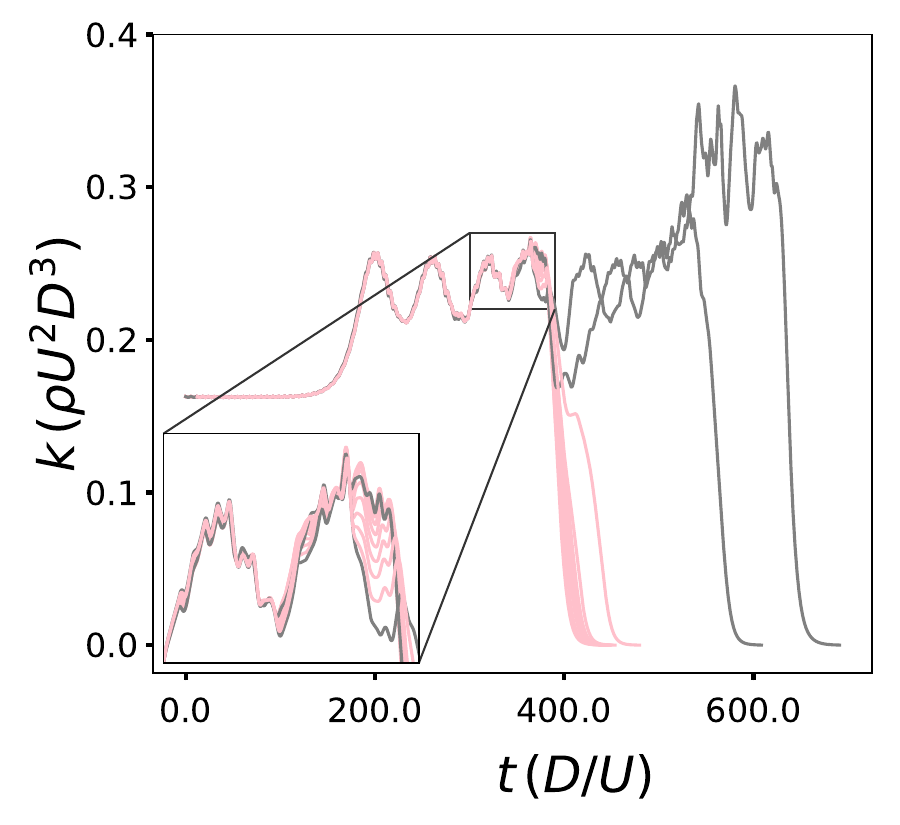}
	\caption{\label{f-manifold}
		(a) (color online) 
		Two-dimensional projection of unstable manifold of LB 
		computed within the Poincar\'e section \refeq{e-pSect} by 
		forward integrating initial points \refeq{e-ICs}. 
		Nearby trajectories that laminarize around the same time are
		colored pink. 
		(b) Time-series of the kinetic energy for the initial 
		conditions \refeq{e-ICs} with
		$\delta \in [0.0, 0.08]$ and 
		$\delta \in [0.97, 0.99]$. The latter shifted by $T$ in time
		to better illustrate the similarity of the time-series. 
		$\delta = 0.08$ and $\delta = 0.97$ are shown
		gray, and the rest,
		which laminarize around the same time, are shown pink.
	} 
\end{figure}

By construction, 
${\bf \tilde{u}}_{LB}$ is a fixed point of the Poincar\'e map. Linear 
stability of ${\bf \tilde{u}}_{LB}$ is then described by the eigenvalue 
problem
\beq
	\left[\Diff \PoincM ({\bf \tilde{u}}_{LB}) \right] \tilde{V}_i = 
	\Lambda_i \tilde{V}_i \,,
\eeq
where $\left[\Diff \PoincM (.) \right]$ is the 
Jacobian of the Poincar\'e map and
$\Lambda_i, \tilde{V}_i$ respectively are the stability eigenvalues and 
eigenvectors of ${\bf \tilde{u}}_{LB}$. We approximate the leading $5$ 
stability eigenvalues and eigenvectors of the LB using Arnoldi 
iteration\rf{Trefethen97} such that 
$\Real \Lambda_1 \geq \Real \Lambda_2 \geq ... \geq \Real \Lambda_5$. 
We found that LB has one unstable direction with the
associated eigenvalue $\Lambda_1 \approx 2.52$ and its leading stable
eigenvalue to be $\Lambda_{2, 3} \approx 0.638 \pm i 0.409$.

In order to approximate the LB's unstable
manifold, we iterate the Poincar\'e map \refeq{e-pMap} starting 
with the initial conditions
\beq
	{\bf \tilde{u}}(\delta) = {\bf \tilde{u}}_{LB} 
							\pm 
							\epsilon \Lambda_1^\delta \tilde{V}_1 \,, 
	\label{e-ICs}
\eeq
where $\epsilon = 10^{-6}$ is a small constant, $\delta \in [0, 1)$ is 
a cyclic 
(${\bf \tilde{u}}(1) \approx \PoincM ({\bf \tilde{u}}(0))$) 
variable that parametrizes the manifold along with the 
discrete time $n$, and $\tilde{V}_1$ is normalized such that 
$||\tilde{V}_1|| = ||{\bf \tilde{u}}_{LB}||$ in $L_2$ (energy)
norm. 
We approximated the unstable manifold of the $LB$ 
by iterating the initial conditions \refeq{e-ICs} using $10$ and $100$  
equally spaced points for $\delta$ respectively in the 
$-\tilde{V}_1$ and $+\tilde{V}_1$ directions.
We visualized this manifold as a projection onto 
$\tilde{V}_1$ and $\Real \tilde{V}_2$ in 
\reffig{f-manifold} (a), where the origin is located at 
${\bf \tilde{u}}_{LB}$. Explicitly, the projections 
are obtained as 
$e_1 = \inprod{{\bf \tilde{u}} - {\bf \tilde{u}}_{LB}}{\tilde{V}_1}$
and
$e_2 = \inprod{{\bf \tilde{u}} - {\bf \tilde{u}}_{LB}}{\Real \tilde{V}_2}$.

\begin{figure}[h!]
	\centering
	(a) \includegraphics[height=0.45\textwidth]{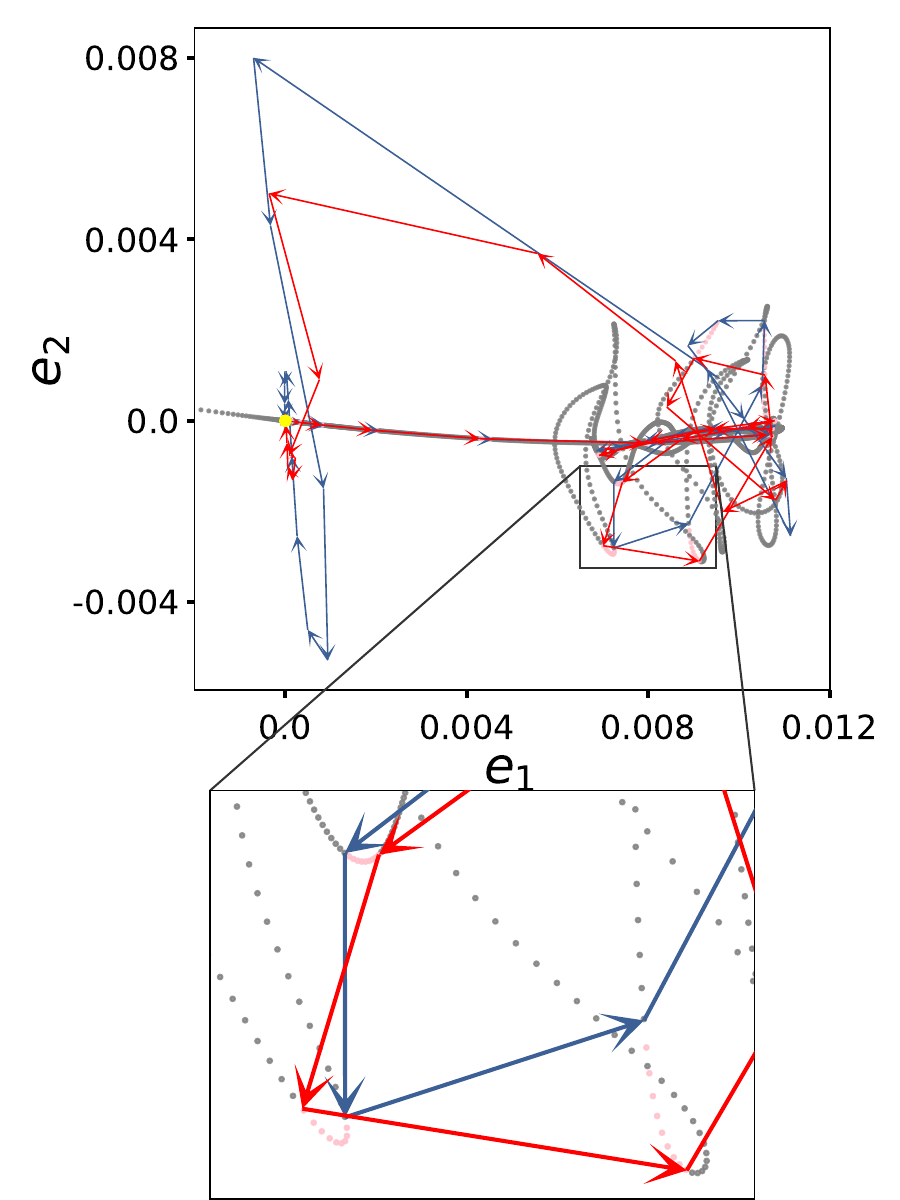} 
	\setlength{\unitlength}{0.45\textwidth}
	(b)
  \begin{picture}(1,0.87722932)%
    \put(0,0){\includegraphics[width=\unitlength,page=1]{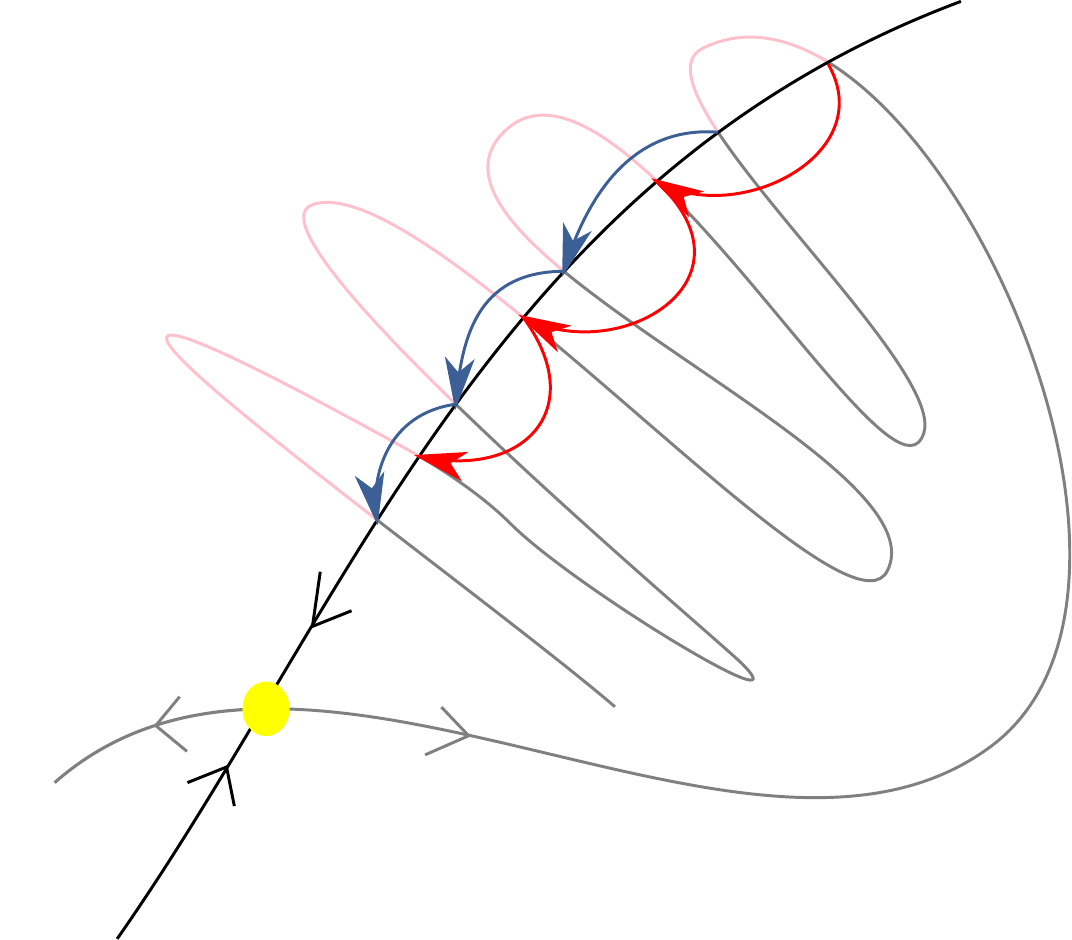}}%
    \put(-0.00214728,0.1446516){\color[rgb]{0,0,0}\makebox(0,0)[lt]{\begin{minipage}{0.11152494\unitlength}\raggedright Laminar\end{minipage}}}%
    \put(0.7,0.56){\color[rgb]{0,0,0}\rotatebox{-39.18865761}{\makebox(0,0)[lt]{\begin{minipage}{0.49171554\unitlength}\raggedright Turbulence\end{minipage}}}}%
    \put(0.39263784,0.13){\color[rgb]{0,0,0}\makebox(0,0)[lt]{\smash{$W_u$}}}%
    \put(0.23,0.09363188){\color[rgb]{0,0,0}\makebox(0,0)[lt]{\smash{$W_s$}}}%
  \end{picture}
	\caption{\label{f-Hclinics}
		(color online) 
		(a) The approximations to the homoclinic orbits (blue and red)
		and the unstable manifold of the LB. 
		Arrows show the direction of time. 
		Inset zooms in to the indicated region in order to better
		visualize that the homoclinic orbits lie inbetween the 
		orbits that laminarize together and those that further 
		explore turbulent parts of the \statesp .
		(b) A ``cartoon'' of the \statesp\ based on the results
		shown in (a). $W_s$ and $W_u$ respectively denote stable 
		and unstable manifolds; the laminarizing trajectories and 
		the homoclinic orbits use the same color scheme as in 
		(a).
		} 
\end{figure}
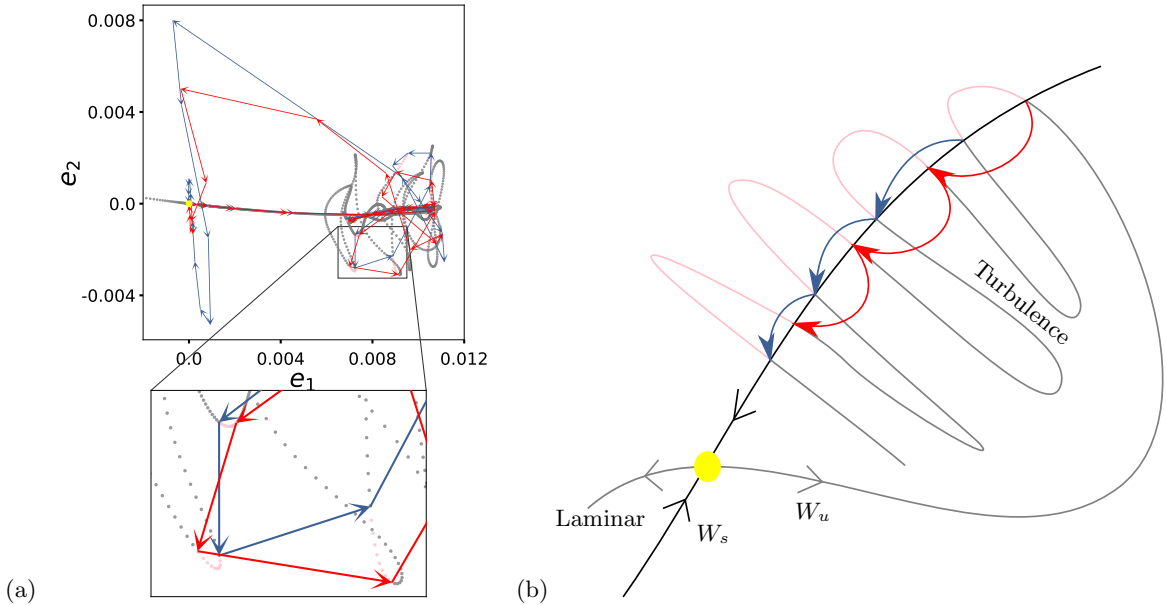

In \reffig{f-manifold}(a), we have shown few points in $-\tilde{V}_1$
since all of these points uneventfully laminarize\footnote{
	The sign, of course, is merely a convention.}. The trajectories, which
are initiated in the $+\tilde{V}_1$ direction, on the other 
hand, explore the chaotic regions of the \statesp\ before eventually 
laminarizing. We observed that the trajectories of the initial 
conditions \refeq{e-ICs} with 
$\delta = 0.0, 0.01, \ldots, 0.07$ and $\delta = 0.98, 0.99$, 
which are colored pink in \reffig{f-manifold}, laminarize around 
the same time when the
time-series for $\delta = 0.98 \mbox{ and } 0.99$ 
is shifted by one-period T; see \reffig{f-manifold}(b). 
This observation suggests that these orbits cross the 
laminar-turbulent boundary, which is set by the stable manifold
of the LB, in the same region of the \statesp. Thus, we expect 
to find \textit{homoclinic orbits} that lie at the transverse
intersections of the 
LB's stable and unstable manifolds within the intervals
$\delta \in (0.07, 0.08)$ and $\delta \in (0.97, 0.98)$.
In order to test this hypothesis, we bisect between the 
trajectories that laminarize after $\zeit = 400 D/U$ and those 
that further visit the turbulent parts of the \statesp\ in both
$\delta$-intervals. This procedure indeed yields orbits that come
back to the LB as can be seen on 
\reffig{f-Hclinics} (a), where we projected these orbits on 
	top of the unstable manifold of \reffig{f-manifold}(a). 
	The projection coordinates are same in both figures.

\begin{figure}
	\centering
	(a) \includegraphics[width=0.45\textwidth]{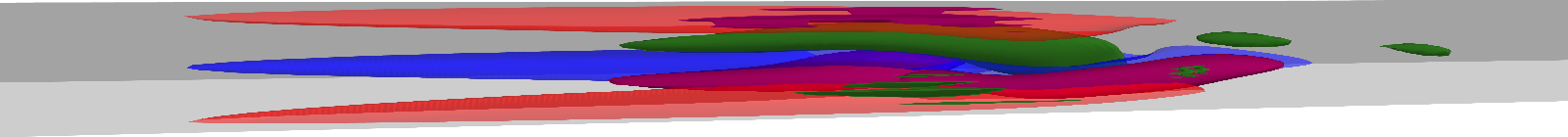} 
	(b) \includegraphics[width=0.45\textwidth]{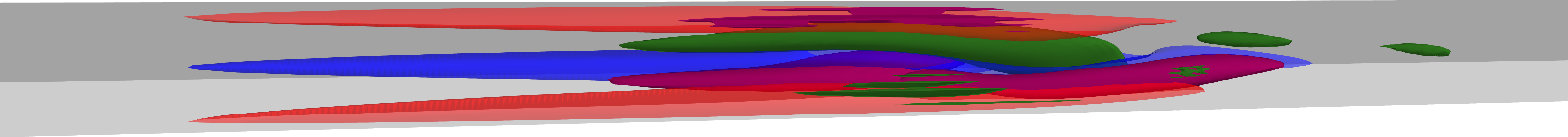} \\
	(c) \includegraphics[width=0.45\textwidth]{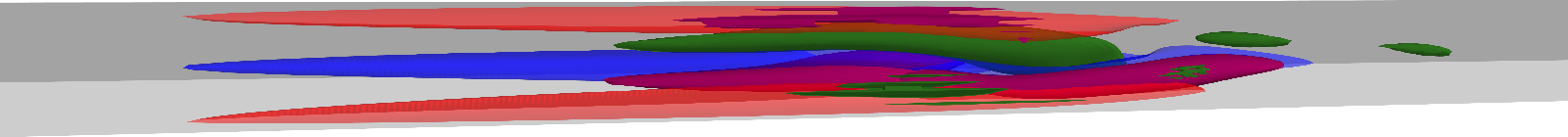} 
	(d) \includegraphics[width=0.45\textwidth]{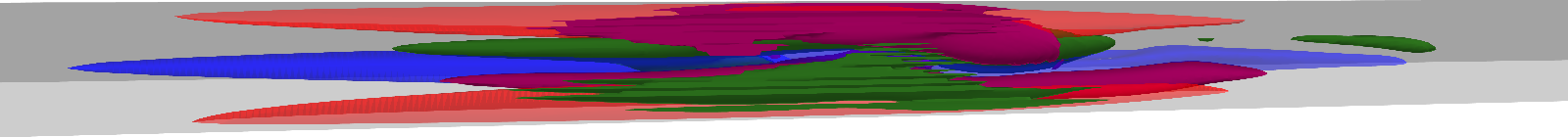} \\
	(e) \includegraphics[width=0.45\textwidth]{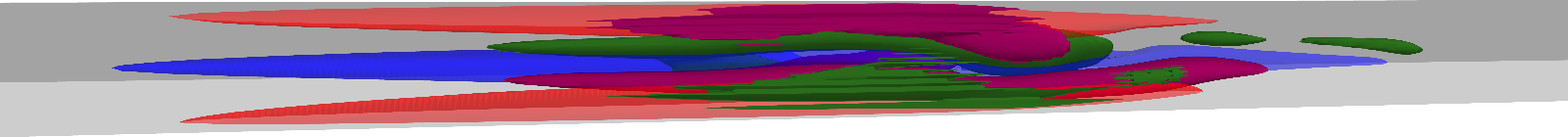} 
	(f) \includegraphics[width=0.45\textwidth]{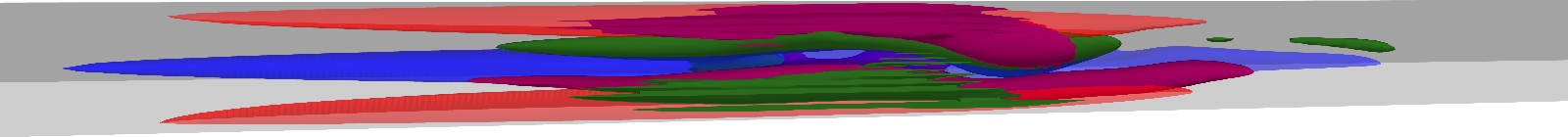} \\
	(g) \includegraphics[width=0.45\textwidth]{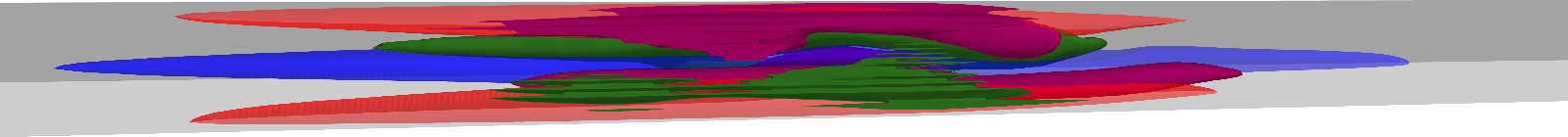} 
	(h) \includegraphics[width=0.45\textwidth]{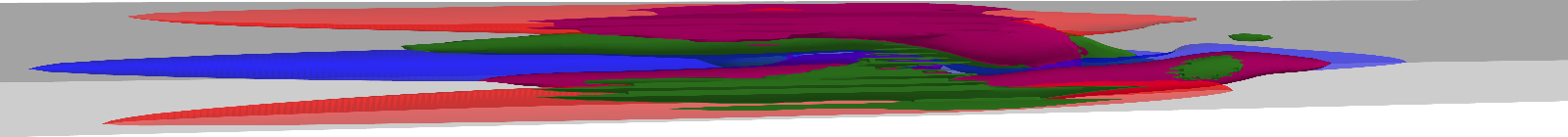} \\
	(i) \includegraphics[width=0.45\textwidth]{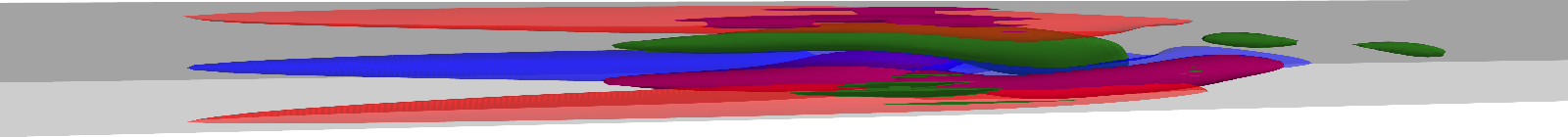} 
	(j) \includegraphics[width=0.45\textwidth]{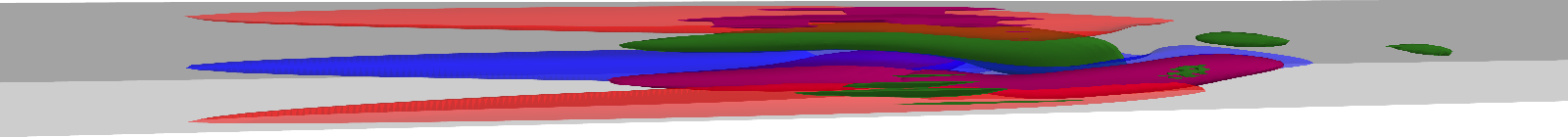} 
	\caption{\label{f-structures} (color online)
		Isosurfaces of the streamwise velocity fluctuations at $\pm 0.3 U$ 
        (red-blue) and the streamwise vorticity at $\pm 0.8 U/D$ (purple-green) 
        visualized for a homoclinic orbit on the Poincar\'e section at times 
        $n = 0, 5, 10, 15, 20, 25, 30, 35, 40, 45$ 
        ($\zeit = 0.0, \ldots, 486.95$) (a--j). 
        A quarter ($\theta \in [0, \pi/2)$) of the domain is shown since the 
        rest can be obtained from the symmetries. 
        Similar visualization 
        of ${\bf \tilde{u}}_{LB}$ is indistinguishable from (a), hence not 
        shown separately.} 
\end{figure}

When the bisections are carried out up to the numerical precision, 
our approximations to the homoclinic orbits come as close as 
$||\tilde{u}_{hc} - \tilde{u}_{LB} || / ||\tilde{u}_{LB}|| \sim 10^{-3}$
to the LB. Further evidence can be seen from the flow structure 
visualizations of \reffig{f-structures}, where the initial and 
final snapshots of homoclinic orbits are virtually indistinguishable. 

The presence of a transverse homoclinic point has profound consequences:

(i) Every forward or backward iteration of a homoclinic point is another 
crossing of the stable and unstable manifolds, hence, 
when the stable and unstable manifolds intersect at one point, they 
do so at infinitely many points as sketched in 
\reffig{f-Hclinics} (b). 

(ii) Transverse intersection of the stable and unstable manifolds 
implies a \textit{horseshoe}\rf{smale}, which necessitates a countable 
infinity of periodic and an
uncountable infinity of aperiodic orbits; hence chaos. 

(iii) In the particular case we consider here, the stable manifold of 
the solution is the boundary of the laminar
equilibrium's domain of attraction, thus, infinitely many 
intersections of the stable and unstable manifolds yield infinitely many 
dynamical routes to puff decay. 

These three points together imply \textit{transient chaos}\rf{hof2006flt}, 
which is
the central assumption of the experiments that study the transition 
to turbulence through a spatiotemporal 
intermittency\rf{AMdABH11,MukHof2018,LSAJAH16}.

Van Veen and Kawahara\rf{VaKa11} computed homoclinic orbits from the
edge of chaos in a small computational cell of the 
plane Couette flow at $\Reynolds = 400$ 
\footnote{\Reynolds\ based on half the velocity difference between 
		  the two walls and half the wall separation.}. 
At this regime, plane Couette flow can be transient in small 
computational cells, however, it is sustained in large-enough 
domains\rf{DSH2010}. Hence they highlighted the resemblance of 
the homoclinic orbits to the so-called turbulent bursting event.
Here, we argue that the 
presence of the homoclinic structure necessitates transient chaos and
thus, provides an explanation of how turbulent puffs can cross
the laminar-turbulent boundary and then rapidly decay to laminar flow.

The symmetry-invariant subspace we studied in this paper 
carries the main features of the pipe flow, however, one should keep
in mind that the complete physics of the pipe flow is richer than 
what we have considered here. We already know that there exists 
a \rpo\ with a threefold azimuthal symmetry
\rf{ChWiKe14} analogous to the one 
we studied here, and it is conceivable that there exists other \rpo s 
at the laminar-turbulent boundary. We speculate that other such 
\rpo s would also have unstable manifolds that contain homoclinic 
orbits, yielding the full complexity of the localized turbulence in 
pipe flow. 

The existence of fractal basin boundaries in transitional shear flows 
was previously suggested in the light of numerical 
experiments\rf{MEF04,RiMeAv16,KES2014}. Our results here clearly show that 
this must indeed be the case for the streamwise-localized turbulence in pipe
flow as a consequence of the homoclinic tangle\rf{MGOY1985}. 

At the first stage of our investigation, we identified a set of trajectories
that belong to a continuum of trajectories that leave the turbulent part of
the \statesp . While not at the focus of the current work, 
we would like to remark that 
this methodology might be exploited for control purposes. In order to 
eliminate turbulence in a transitional setting like ours, the flow must be 
driven towards the laminarizing side of the edge state's stable manifold. 
However, there exists no method for approximating this manifold since the 
time-backward integration is not possible. 
Therefore, locating parts of the unstable manifold that intersect the stable
manifold could be a starting point for perturbing the system in this 
direction.

\edit{
We confirmed the 
existence of homoclinic orbits at higher resolutions by partially 
repeating our computations with $(K, M, N ) = (383, 47, 96)$ and half 
of the time-step. We would like to note that increasing the 
resolution changes the appearance of the computed manifold and precise 
location of homoclinic orbits. We believe that these variations are due to 
the differences in the flow regime upon changing the resolution. 
Avila \etal\rf{AvMeRoHo13} already reported that the changes in resolution 
shifted the bifurcation points while leaving the qualitative scenario 
unchanged and we think that our situation here is similar. 
Our result, along with the recent paper by Lustro \etal \rf{LKVVSK2019}, who
followed a very similar methodology to illucidate the homoclinic tangency in 
a minimal plane Couette flow, suggest numerical methods for computing 
homoclinic orbits in high-dimensional systems as an important technical 
challenge. 
}

In summary, we have studied the unstable manifold of a
streamwise-localized \rpo\ on the laminar-turbulent boundary of 
pipe flow and showed that it accommodates homoclinic orbits. 
We argue that this structure \edit{qualitatively} explains the transient chaos
that is observed in the computer and laboratory experiments of the
spatially localized turbulence in pipe flow. 

\acknowledgements

ASD acknowledges the financial support from OeAD Sonderstipendien,
\"{O}sterreichische Austauschdienst-Gesellschaft -- Austria 
and 
Dean's Undergraduate Research Fund Grant (DURF),
College of Arts and Sciences (NYU). 
We are also grateful to the Scientific Computing Unit at IST Austria,
which maintains the IST high performance computing cluster, where all 
simulations presented in this paper were performed.



\bibliography{../bibtex/neubauten}

\end{document}